# The Role of Cloud-MANET Framework in the Internet of Things (IoT)

Tanweer Alam [✉], Mohamed Benaida
Faculty of Computer and Information Systems
Islamic University of Madinah, Madinah, Saudi Arabia
tanweer03@iu.edu.sa, md.benaida@gmail.com

**Abstract**—In the next generation of computing, Mobile ad-hoc network (MANET) will play a very important role in the Internet of Things (IoT). The MANET is a kind of wireless networks that are self-organizing and auto connected in a decentralized system. Every device in MANET can be moved freely from one location to another in any direction. They can create a network with their neighbors' smart devices and forward data to another device. The IoT-Cloud-MANET framework of smart devices is composed of IoT, cloud computing, and MANET. This framework can access and deliver cloud services to the MANET users through their smart devices in the IoT framework where all computations, data handling, and resource management are performed. The smart devices can move from one location to another within the range of the MANET network. Various MANETs can connect to the same cloud, they can use cloud service in a real time. For connecting the smart device of MANET to cloud needs integration with mobile apps. My main contribution in this research links a new methodology for providing secure communication on the internet of smart devices using MANET Concept in 5G. The research methodology uses the correct and efficient simulation of the desired study and can be implemented in a framework of the Internet of Things in 5G.

## 1 Introduction

This research is a step forward in the field of MANET, cloud, and IoT in 5G heterogeneous networks where I propose a new framework using cloud computing for communicating in the MANET and internet of smart devices of the 5G network. The IoT can be described as "a pervasive and ubiquitous system which empowers screening furthermore control the physical earth by collecting, processing, also analyzing that information created eventually sensors" [1]. The proposed research work in this study is an enhancement and implementation of existing mobile ad hoc network communication using the cloud in the framework of the IoT. The research outcome is to establish a new framework for communication. The proposed research uses the correct and efficient simulation of the desired study and can be implemented in a framework of the Internet of Things. The most wireless network of today consists of cells. Each cell contains a base station that can be wired or wirelessly connected [2]. The smart devices have a very useful feature of Wi-Fi Direct [3]. Using this feature any device can connect to each other and form an ad hoc network [4]. If one device has internet then this device can connect to the cloud and create an internet of smart devices network. It is expected that by 2020, the development of the internet of smart devices connected together exponentially with 50 billion smart devices [5]. This development will not depend on mankind's population but the reality that units we utilize consistently (See fig 1 for statistics between 2003-2020) [6]. The reality of interconnectedness things is cooperating man to machines and machine to another machine. They will be talking to each other [7]. But Monitoring and tracking of movable devices are one of the most comprehensive issues [8].

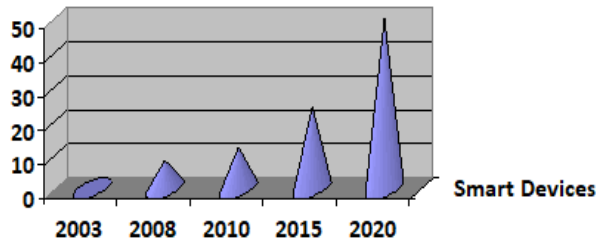

Fig. 1. IOT device statistics between 2003-2020 [43]

This evolutionary paradigm enables its users to deploy a connection to a network of computing resources in an effortless fashion, where users can rapidly scale up or down their demands with trivial interaction from the service provider [9].

### 1.1 The growth of the Internet of Things

The growth of the internet of things initially started in 2008 by connecting the physical objects to the internet [10]. The physical objects are connected to a smart database that has a collection of smart data. The framework has the image recognition technology for identifying the physical object, buildings, peoples, logo, location etc. for business and customers [11]. Now the internet of things is shifting from information-based technology to operational based technology i.e. IPV4 (man 2 machines) to IPV6 (machine 2 machines) [12]. It combines sensors, smart devices and interfaces like Smart Grid [13]. In a wider respect, each of the previous consumers has their concerns over cloud computing vulnerabilities and challenges which might prevent them from their objectives. The components of IoT are Identifiers, Sensors, Communications, Computations, Services, and Semantics [14]. The ubiquitous computation that has the capacity of intelligent physical objects that execute on the computation framework [15]. Internet Protocol (IPV6) using ubiquitous computing that covers the area of network and support talking from machine to machine [16]. IPv4 internet has a drawback to adding billions of smart gadgets together, but it is possible in IPv6 internet because it enables internet of things to connect billions of smart gadgets together securely [17].

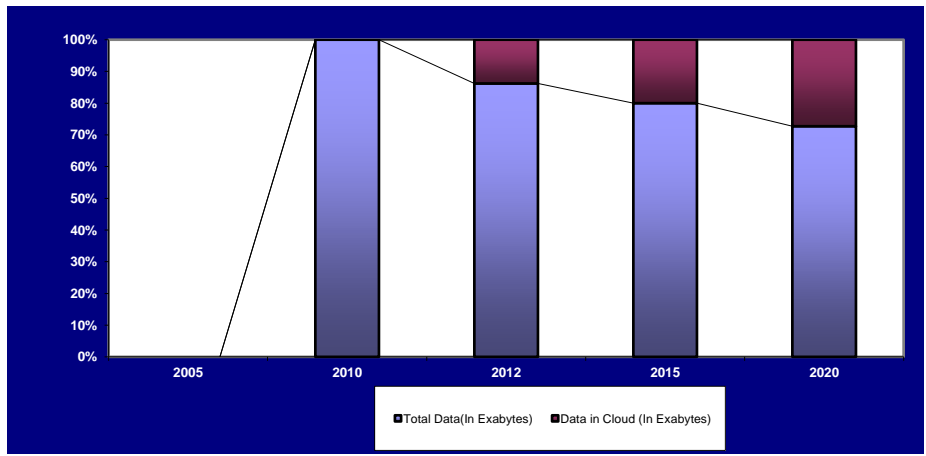

Fig. 2. Siemens Analysis of the growth of cloud-based data vs. total data [17]

Connection using ubiquitous computing that uses the fixed cell network or mobility by using sensor connectivity [18]. These technologies should be enhanced regularly so that it allows the progress of internet of smart devices including multi-sensor framework to store, compute, analyze and process capabilities with smaller in size and lowest energies required [19]. The main contribution of this article links a new secure communication model using cloud computing and MANET technologies in the area of the internet of things. The communication security idea depends on three main points in the designing of the internet of things architecture [20]. It is not easy to manage information getting from millions of sensors in a centralized framework of smart devices collection [21]. It is not easy to manage network resources in a large network that can collect environment information from the centralized framework [22]. It is very hard to manage sensors that execute the same kind of data parallel and stored in the centralized framework. According to Siemens research, up to 2020, near about 26 billion physical objects will be connected together on the internet (See fig 2) [23]. That time is not far away when billions of physical things linked together in real time. They can communicate with each other and forwarding and process required data in the cloud. But there is a lack of technical standardization security perspective on the internet of smart thing.

### 1.2 Cloud Computing

Cloud computing has been regarded as one of the most popularized computing paradigms [24]. It came likewise an outcome for developments done past computing paradigms which incorporate parallel computing, grid computing, disseminated computing also other computing paradigms [25]. Cloud computing gives its customers three essential administration models: SaaS, PaaS, and IaaS. Software as a service (SaaS) is mainly intended to end users who need to use the software as a part of their daily activities [26]. Platform as a service (PaaS) is mainly intended for application developers who need platforms to develop their software or application [27]. Infrastructure as a service (IaaS) is mainly intended to network architects who need infrastructure capabilities [28].

### 1.3 Cloud-MANET Model

The smart device to smart device communication in the cloud-MANET framework of the internet of things is a novel methodology that discovers and connected nearby smart devices with no centralized infrastructure [29]. The proposed technique will be very useful in machine to machine (M2M) networks because, in the M2M network, there are several devices nearby to each other. The smart device users will use cloud service to discover the devices, minimize useful information in big data and can process videos, images, text, and audio. In the proposed framework, the smart device will consider as service nodes.

### 1.4 Motivation

The MANET is a very popular network to get connected anywhere at any time [30]. Cloud provides service for storing and accessing information. The integration of cloud and MANET provides the facilities to access the cloud inside MANET of smart devices. In real life situation, the group of smart device users wants to connect to each other in a meeting at a place where no network services are present. These users may form MANET among smart devices. Also, they can use cloud service only if one device has internet in the group.

### 1.5 Research organization

The organization of the rest of the research paper is as follows: 1- Introduction, 2- Cloud-MANET in IoT, 3-Communication security issues, 4- Working with proposed research and 5-represents the conclusion.

### 1.6 Research contributions

This research paper proposes the Cloud-MANET-IoT framework for communication among smart devices. The following procedure is followed in the proposed research. 1. Form a MANET, 2. Access the ad hoc network in the range of network, 3. Register smart devices in MANET, 4. Register MANET smart devices in Cloud, 4. Implement IoT-Cloud-MANET model among all smart devices and 5. Start communication. The smart device to smart device communication in the IoT-MANET framework is a novel methodology that discovers and connected nearby smart devices with no centralized infrastructure. The existing cellular network doesn't allow to connect all smart devices without centralized infrastructure even if they are very near to each other. The proposed technique will be very useful in machine to machine (M2M) networks because, in the M2M network, there are several devices nearby to each other. So, the implementation of MANET model in the smart device to smart device communication can be very efficient and useful to save power as well as the efficiency of spectrums. The cloud-based services in MANET modeling for the device to device communication can be a very useful approach to enhance the capabilities of smart devices. The smart device users will use cloud service to discover the devices, minimize useful information in big data and can process videos, images, text, and audio.

## 2 Cloud-MANET in IoT

### 2.1 Discovering the smart devices in MANET

The Ad hoc network can connect all smart devices in the decentralized system [31]. The smart devices are located on the 3D plane in the directions of the x-axis, y-axis, and z-axis. The whole area is divided into cells over the wireless network. The area of all cells is fixed so that the smart devices can travel within the range of MANET Cell [32]. The smart device discovers the neighborhood devices in a binary digit within the same cell area. For discovering the smart devices, the hidden Markov model (HMM) is utilized in the 2Dimensional plane area. This model is connected to the working area and devices move inside the area and this model found neighborhood devices within the range [33]. We form the transition matrix in the area of the wireless network, discover all the smart devices and put in the transition matrix. The following parameters are used for discovering the smart devices. This model contains the following parameters. Let S=$S_1$, $S_2$………$S_N$ where S=state, $S_1$ is the first state, $S_2$ is the second state and so on [34].

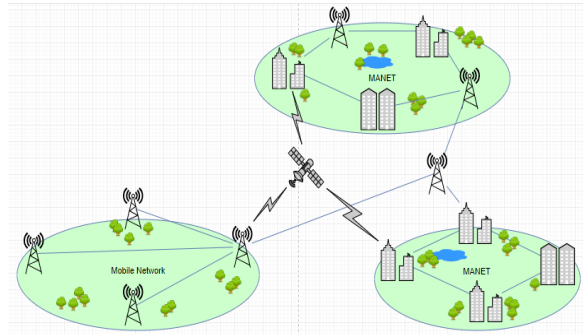

Fig. 3: MANET

Each cell depends on one state. The transition matrix probability P= $P_{ij}(1≤i≤N)$ where $P_{ij}$ characterized to move likelihood from $S_i$ to $S_j$. Geometrically, $P_{ij}$ is just significant if Si, $S_j$ are neighborhood states. Now rearrange states to move up, down, left and right [35]. Whatever is left components within the framework are all 0s. The following figure represents the transition matrix using a hidden Markov model.

Fig. 4: Discovering area by HMM model

The smart devices in MANET can be used to discover the signals using the Viterbi algorithm. Let $O_1$, $O_2$,. . . , $O_n$ are the observation of discovering the devices. Every smart device sends a report of observations in meanwhile. This algorithm discovers the way at each step by maximizing the likelihood. This process is so expensive and time consuming for the rush of devices. The matrix represents the information in every cell. When the smart device enters in a new cell, it removes previous data and updates the information with the new data. For discovering the smart devices, the gradient model (GM) works to find the devices and share the ideas to development and send the information. GM finds the gradient distribution over the time. If time=0 then the gradient value will be 1. If time is greater than the total time then the value of the gradient will be 0. Otherwise, the gradient is proportional to one upon time power of the event.

Table 1: PBM, HMM vz. GM statistics

|  | PBM | HMM | GM |
|---|---|---|---|
| Average Path Length | 8 | 8.3 | 15 |
| Average Stretch Factor | 2.2 | 2.3 | 3.91 |
| Success Rate | 96.5% | 95% | 89.5% |

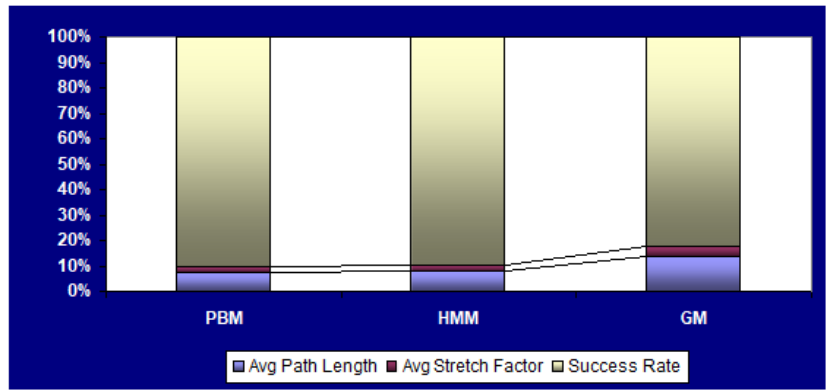

Fig. 5: PBM, HMM and gradient Model statistics

So, the achievement rate of the PBM model is best in the examination of HMM and Gradient Model. So, we utilize the PBM model for outlining the Ad Hoc Network among smart devices.

### 2.2  Place smart devices in the range of MANET

The smart devices are put within range of MANET that considers coverage and connectivity. Every smart device is expected to have a connection and an altered correspondence range. At first, MANET has a dynamic state. MANET is inactive all the time, when one device wants to make the connection with another device then it creates the connection with their neighborhood within the discovering area. A resting smart device additionally occasionally awakens to enter into a ready state. The ready state detects the devices and comes back to the dynamic state.

### 2.3 Implement MANET

The smart device to smart device communication in the IoT-MANET framework is a novel methodology that discovers and connected nearby smart devices with no centralized infrastructure. The existing cellular network doesn't allow connecting all smart devices without centralized infrastructure even if they are very near to each other. The proposed technique will be very useful in machine to machine (M2M) networks because, in the M2M network, there are several devices nearby to each other. So, the implementation of MANET model in the smart device to smart device communication can be very efficient and useful to save power as well as the efficiency of spectrums. The cloud-based services in MANET modeling for the device to device communication can be a very useful approach to enhance the capabilities of smart devices [36]. The smart device users will use cloud service to discover the devices, minimize useful information in big data and can process videos, images, text, and audio. In this article, we proposed a new framework to enhance the capability of MANET and cloud computing on the internet of smart devices that can be useful in the 5G heterogeneous network. In the proposed framework, the smart device will consider as service nodes. This model also covers the security and reliability as well as vulnerability issue of communication. The smart devices are electronic devices that are connected to other devices or network through network protocols such as a smartphone, tablets, smartwatch etc. The cloud computing enables sharing the resources, storages, and services using mobile applications. This study is focused on secure communication among smart devices in the area of Cloud-MANET. In Cloud-MANET, the smart devices are dynamically joined and created a network on their own called MANET and they can access cloud service [37]. This algorithm is implemented as a mobile application and tested in cloud-MANET of smart devices. The results are found positive and can be implemented in the framework of the internet of things in 5G heterogeneous networks.

The Mobile ad hoc network is a kind of wireless networks that are self-organizing and auto connected in a decentralized system. Every node in MANET can be moved freely from one location to another in any direction [38]. They can create a network with their neighbors' smart devices and forward data to another device like a router. The cloud-MANET framework of smart devices is composed of cloud computing and MANET [39]. This framework can access and deliver cloud services to the MANET users through their smart devices where all computations, data handling, and resource management are performed. The smart devices can move from one location to another. Various MANETs can connect to the same cloud, they can use cloud service in a real time. For connecting the smart device of MANET to cloud needs integration with mobile apps. The MANET model of smart devices in a local communication can work very well using the cloud, it is failed when it connects in exist wired networking framework. For working in wireless and wired infrastructure, an access point will be required like gateways. The communication process to connect a smart device with another device in the cloud-MANET framework, every smart device must be configured universally using routing IP address [39]. Also, every device requires searching neighbor device as well as gateways that use its prefixed and assign the universally routing IP addresses. When these devices will connect to each other using cloud services then the main issue is the security of communication. The question is arising here how is the communication secure in the public cloud and MANET model? The proposed algorithm for secure communication in the IoT-MANET model is implemented and integrated with mobile apps. Java programming is used to develop a mobile app. This mobile app should be installed on every smart device of MANET infrastructure. The experiment was conducted using the smart device with proposed research implementation and other existing algorithms were installed on same configuration devices. We notice the performance of the proposed research was better than all existing algorithms.

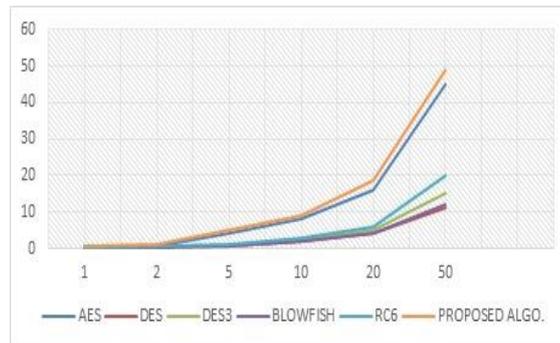

Fig. 6. Algorithms performance in comparisons of block size and time is taken.

In the future, this algorithm may be integrated with Wi-Fi Direct protocol and implement it on smart devices. The experimental results of the proposed algorithm have a better performance than all other existing algorithms. DES algorithm showed the poor performance while RC6 showed very good performance. This study can play very important roles in the framework of the Internet of Things where smart devices will communicate with each other securely.

## 3 Working with proposed research

The IoT-MANET framework is an integrated model of IoT, Cloud computing, and MANET technologies. The functionality of MANET is depended on the mobility of its nodes and connectivity also resources such as storage and energy efficiency [40]. In Cloud computing, cloud providers retain network infrastructure, storage facilities, and software applications that support flexibility, efficiency, and scalability [41]. In Cloud MANET mobility model, smart devices of MANET can communicate with each other but at least one smart device must be connected to cellular or Wi-Fi networks. All smart devices of MANET should be registered in cloud individually. The proposed model will activate in disconnected mode. When a MANET is activated then cloud services will activate in a real-time and provide services to the smart devices of MANET [42]. The smart devices send a request to the cloud for a session of connectivity. Cloud provides the best connection to the smart device. The expression in the integral will be 0 if the limit tends to ∞. After computing session life by using the above probabilistic function, every smart device requires computing the values of σ and μ. These two parameters are related to the connection establishment among MANETs and Cloud service that can be measured through smart devices using the following function. eμ(1/2)σ². When a smart device estimates the connection life between MANET and Cloud, it will transfer or receive data securely. The connection will be activated and stability will be high. We consider that every smart device is assured to establish the route between MANET and cloud when they create a session in the cloud. The smart devices can move through the maximum speed 20m/s from one location to another location by using the Gauss-Markov mobility model. The proposed framework consists of three algorithms. The firstly discover the smart devices within the range of MANET then discovering the gateway point to connect to the cloud and the thirdly is used to establish a connection, provide session and transfer data from one smart device to another using cloud as a service.

Table 2: Transmission in Cloud-MANET of Smart devices at 10 m/s.

| Smart Devices | t =0.1 | t =0.2 | t=0.4 | t =0.6 | t=0.8 | t =1 |
|---|---|---|---|---|---|---|
| 5 | 2.1 | 2.2 | 2.3 | 2.2 | 2.2 | 2.3 |
| 10 | 3.1 | 3.3 | 3.3 | 3.3 | 3.4 | 3.3 |
| 50 | 4.9 | 4.8 | 4.7 | 4.6 | 4.8 | 4.7 |

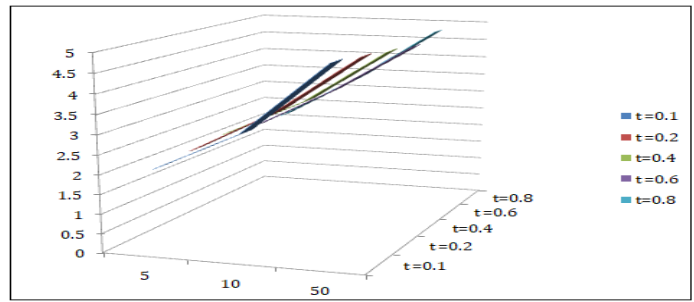

Fig. 7. Transmission in Cloud-MANET of Smart devices at 10 m/s.

Table 3: Transmission in Cloud-MANET of Smart devices at 20 m/s.

| Devices | t =0.1 | t =0.2 | t =0.4 | t =0.6 | t =0.8 | t =1 |
|---|---|---|---|---|---|---|
| 5 | 2 | 2.1 | 2 | 2.2 | 2.5 | 2.4 |
| 10 | 3 | 3.2 | 3.1 | 3.5 | 3.4 | 3.3 |
| 50 | 7 | 7.2 | 7.5 | 7.8 | 7.4 | 7.3 |

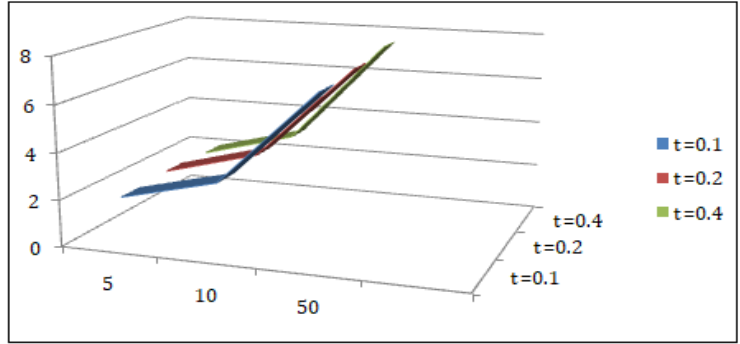

Fig. 8. Transmission in Cloud-MANET of Smart devices at 20 m/s.

Table 4: Transmission in Cloud-MANET of Smart devices at 50 m/s.

| Devices | t =0.1 | t =0.2 | t =0.4 | t =0.6 | t =0.8 | t =1 |
|---|---|---|---|---|---|---|
| 5 | 2.1 | 1.9 | 1.9 | 2.2 | 2.3 | 2.42 |
| 10 | 3 | 2.9 | 3.1 | 2.8 | 2.7 | 3 |
| 50 | 5.1 | 5.5 | 7.5 | 5.3 | 5.2 | 5.6 |

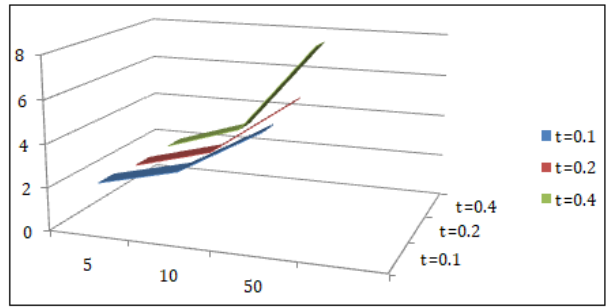

Fig. 9. Transmission in Cloud-MANET of Smart devices at 50 m/s.

The following algorithm is used to find the new position of the smart devices in MANET.
Current Location:- Location L1=getNewLocation (new     Point(x1,y1));
New location:- Location L2=getNewLocation (new  Point(x2,y2));
Find distance between L1 and L2:- Distance $(d)=\sqrt{(x2-x1)2-(y2-y1)2}$
Find random location (X, Y) of smart device at the diagonal of triangle.
    X=Math.random(d.getX());

```
        Y=Math.random(d.getY());
```
Find the actual location of the smart device according to the diagonal of the triangle. The device may be up or down from diagonal. If the device is upper than the diagonal then increase the value of X and Y as follows.
X= X + ΔX;
Y= Y + ΔY ;
Otherwise
X= X − ΔX;
Y= Y − ΔY;
Return the new Location(X, Y)

The following algorithm will compute the velocity of smart devices in the Cloud-MANET framework.
Input: Number of smart devices, transmission function value (t).
Output: Velocities of smart devices.
Initialization: Counter=0, V=0.
Step 1: Find all the probabilities of smart devices in each direction.
Step 2: Find entropy per symbol.
Step 3: Find Transmission in bits/sec.
Step 4: Find velocity (V) in m/s.
Step 5: Counter=Counter+1;
if counter < Number of devices go to step 2 otherwise stop.

The complexity of the algorithm is $O(n^2)$.

The Proposed mobility model had been implemented using two mobile applications. These mobile applications are verified on three Samsung devices. One of them is supported by the 4G network and another two are supported by 3G networks. Amazon Web Services (AWS) are used for implementing cloud services. This cloud service will connect to the MANETs. The mobile apps should be installed on every smart device. After installing mobile apps for the devices, the device should be registered in the cloud. The cloud will generate a device id for every smart device that is registered to the cloud. A smart device can communicate with another smart device within the range of the same MANET or another MANET using cloud services. The smart user will be activated by mobile apps. When he opened smart apps then he had connected with Amazon cloud service automatically and start to communicate with another device. The Amazon cloud provides relational database services for storing smart device information, requests, communicated messages, and neighborhood smart devices information.

## 4 Conclusion

This study enhances the role of a Cloud-MANET framework for communication among internet of smart devices. The research outcome is to establish a new framework in IoT. The proposed research is used the correct and efficient simulation of the desired study and can be implemented in a framework of the Internet of Things. The cloud-based services in MANET modeling for the device to device communication can be a very useful approach to enhance the capabilities of smart devices. The smart device users will use cloud service to discover the devices, minimize useful information in big data and can process videos, images, text, and audio. The proposed algorithm for the communication in the IoT-MANET model is implemented and integrated with mobile apps. Java programming is used to develop such mobile apps.

## 5      References

[1] Tanweer Alam, "A Reliable Communication Framework and Its Use in Internet of Things (IoT)", International Journal of Scientific Research in Computer Science, Engineering and Information Technology (IJSRCSEIT), ISSN: 2456-3307, Volume 3, Issue 5, pp.450- 456, May-June.2018 URL: http://ijsrcseit.com/CSEIT1835111

[2] Nasser, N., Hasswa, A. and Hassanein, H., 2006. Handoffs in fourth generation heterogeneous networks. *IEEE Communications Magazine*, *44*(10), pp.96-103.

[3] Want, R., Schilit, B.N. and Jenson, S., 2015. Enabling the internet of things. *Computer*, *48*(1), pp.28-35.

[4] Pelusi, L., Passarella, A. and Conti, M., 2006. Opportunistic networking: data forwarding in disconnected mobile ad hoc networks. *IEEE Communications Magazine*, *44*(11).

[5] https://www.cisco.com/c/en/us/solutions/collateral/enterprise/cisco-on-cisco/Cisco_IT_Trends_IoE_Is_the_New_Economy.html

[6] Statista. Internet of Things (IoT): number of connected devices worldwide from 2012 to 2020, [online], URL =http://www.statista.com/statistics/471264/iot-number-of-connected-devices-worldwide, 2016.

[7] T. Alam and M. Aljohani, "An approach to secure communication in mobile ad-hoc networks of Android devices," 2015 International Conference on Intelligent Informatics and Biomedical Sciences (ICIIBMS), Okinawa, 2015, pp. 371-375. doi: 10.1109/ICIIBMS.2015.7439466

[8] T. Alam and M. Aljohani, "Design and implementation of an Ad Hoc Network among Android smart devices," 2015 International Conference on Green Computing and Internet of Things (ICGCIoT), Noida, 2015, pp. 1322-1327. doi: 10.1109/ICGCIoT.2015.7380671

[9] M. Aljohani and T. Alam, "An algorithm for accessing traffic database using wireless technologies," 2015 IEEE International Conference on Computational Intelligence and Computing Research (ICCIC), Madurai, 2015, pp. 1-4. doi: 10.1109/ICCIC.2015.7435818

[10] M. Aljohani and T. Alam, "Design an M-learning framework for smart learning in ad hoc network of Android devices," 2015 IEEE International Conference on Computational Intelligence and Computing Research (ICCIC), Madurai, 2015, pp. 1- 5. doi: 10.1109/ICCIC.2015.7435817

[11] Tanweer Alam and Mohammed Aljohani. 2016. Design a New Middleware for Communication in Ad Hoc Network of Android Smart Devices. In Proceedings of the Second International Conference on Information and Communication Technology for Competitive Strategies (ICTCS '16). ACM, New York, NY, USA, Article 38, 6 pages. DOI: http://dx.doi.org/10.1145/2905055.2905244

[12] Aljohani, Mohammed, and Tanweer Alam. "Real Time Face Detection in Ad Hoc Network of Android Smart Devices." Advances in Computational Intelligence: Proceedings of International Conference on Computational Intelligence 2015. Springer Singapore, 2017.DOI: https://doi.org/10.1007/978- 981-10-2525-9_24

[13] Alam, Tanweer. "Middleware Implementation in Cloud-MANET Mobility Model for Internet of Smart Devices", International Journal of Computer Science and Network Security, 17(5), 2017. Pp. 86-94.

[14] Sharma, Abhilash, Tanweer Alam, and Dimpi Srivastava. "Ad Hoc Network Architecture Based on Mobile Ipv6 Development." Advances in Computer Vision and Information Technology (2008): 224.

[15] Alam, Tanweer. "Fuzzy control based mobility framework for evaluating mobility models in MANET of smart devices." ARPN Journal of Engineering and Applied Sciences 12.15 (2017): 4526-4538.

[16] Singh, Parbhakar, Parveen Kumar, and Tanweer Alam. "Generating Different Mobility Scenarios in Ad Hoc Networks.", International Journal of Electronics Communication and Computer Technology, 4(2), 2014

[17] Zanella, A., Bui, N., Castellani, A., Vangelista, L. and Zorzi, M., 2014. Internet of things for smart cities. *IEEE Internet of Things journal*, *1*(1), pp.22-32.

[18] Estrin, D., Culler, D., Pister, K. and Sukhatme, G., 2002. Connecting the physical world with pervasive networks. *IEEE pervasive computing*, *1*(1), pp.59-69.

[19] Botta, A., De Donato, W., Persico, V. and Pescapé, A., 2016. Integration of cloud computing and internet of things: a survey. *Future Generation Computer Systems*, *56*, pp.684-700.

[20] Atzori, L., Iera, A. and Morabito, G., 2010. The internet of things: A survey. *Computer networks*, *54*(15), pp.2787-2805.

[21] Akyildiz, I.F., Su, W., Sankarasubramaniam, Y. and Cayirci, E., 2002. Wireless sensor networks: a survey. *Computer networks*, *38*(4), pp.393-422.

[22] D. Chakraborty, A. Joshi, Y. Yesha, and T. Finin. Toward distributed service discovery in pervasive computing environments. IEEE Transactions on Mobile Computing, 5(2):97–112, Feb 2006.

[23] https://www.siemens.com/innovation/en/home/pictures-of-the-future/digitalization-and-software/internet-of-things-facts-and-forecasts.html

[24] Zhang, Q., Cheng, L. and Boutaba, R., 2010. Cloud computing: state-of-the-art and research challenges. *Journal of internet services and applications*, *1*(1), pp.7-18.

[25] Alam, Tanweer, and B. K. Sharma. "A New Optimistic Mobility Model for Mobile Ad Hoc Networks." International Journal of Computer Applications 8.3 (2010): 1-4. doi: 10.5120/1196-1687

[26] Alam, Tanweer, and Tapesh Kumar Tyagi. "A Random Waypoint Model for Mobility in Ad Hoc Network Simulation Area." Journal of Advanced Research in Computer Engineering 3.1 (2009): 13-17.

[27] Alam, Tanweer, et al. "Scanning the Node Using Modified Column Mobility Model." Computer Vision and Information Technology: Advances and Applications (2010): 455.

[28] Alam, Tanweer. " A reliable framework for communication in internet of smart devices using IEEE 802.15.4." ARPN Journal of Engineering and Applied Sciences 13.10 (2018)

[29] Alam, Tanweer, Parveen Kumar, and Prabhakar Singh. "SEARCHING MOBILE NODES USING MODIFIED COLUMN MOBILITY MODEL.", International Journal of Computer Science and Mobile Computing, (2014).

[30] Pelusi, L., Passarella, A. and Conti, M., 2006. Opportunistic networking: data forwarding in disconnected mobile ad hoc networks. IEEE communications Magazine, 44(11).

[31] Conti, M. and Giordano, S., 2007. Multihop ad hoc networking: The reality. IEEE Communications Magazine, 45(4).
[32] Li, F. and Wang, Y., 2007. Routing in vehicular ad hoc networks: A survey. IEEE Vehicular technology magazine, 2(2).
[33] Sparacino, F., Wren, C., Azarbayejani, A. and Pentland, A., 2002, June. Browsing 3-D spaces with 3-D vision: body-driven navigation through the Internet city. In null (p. 224). IEEE.
[34] Gallager, R.G., 1996. Markov Processes with Countable State Spaces. In Discrete Stochastic Processes (pp. 187-222). Springer, Boston, MA.
[35] Tierney, L., 1994. Markov chains for exploring posterior distributions. the Annals of Statistics, pp.1701-1728.
[36] Huang, D., Zhang, X., Kang, M. and Luo, J., 2010, June. MobiCloud: building secure cloud framework for mobile computing and communication. In 2010 fifth IEEE international symposium on service oriented system engineering (pp. 27-34). Ieee.
[37] Bonomi, F., Milito, R., Zhu, J. and Addepalli, S., 2012, August. Fog computing and its role in the internet of things. In Proceedings of the first edition of the MCC workshop on Mobile cloud computing (pp. 13-16). ACM.
[38] Bai, F. and Helmy, A., 2004. A survey of mobility models. *Wireless Adhoc Networks. University of Southern California, USA*, *206*, p.147.
[39] Zhou, B., Dastjerdi, A.V., Calheiros, R.N., Srirama, S.N. and Buyya, R., 2015, June. A context sensitive offloading scheme for mobile cloud computing service. In *Cloud Computing (CLOUD), 2015 IEEE 8th International Conference on* (pp. 869-876). IEEE.
[40] Angin, P., 2013. *Autonomous Agents-Based Mobile-Cloud Computing* (Doctoral dissertation, Purdue University).
[41] Dantu, K., Rahimi, M., Shah, H., Babel, S., Dhariwal, A. and Sukhatme, G.S., 2005, April. Robomote: enabling mobility in sensor networks. In *Proceedings of the 4th international symposium on Information processing in sensor networks* (p. 55). IEEE Press.
[42] Bhardwaj, S., Jain, L. and Jain, S., 2010. Cloud computing: A study of infrastructure as a service (IAAS). *International Journal of engineering and information Technology*, *2*(1), pp.60-63.
[43] Chiang, M. and Zhang, T., 2016. Fog and IoT: An overview of research opportunities. *IEEE Internet of Things Journal*, *3*(6), pp.854-864.
[44] Tanweer Alam and Mohamed Benaida, "CICS: Cloud–Internet Communication Security Framework for the Internet of Smart Devices", International Journal of Interactive Mobile Technologies, 2018